\documentclass[10pt,prl,twocolumn,natbib,showkeys] {revtex4}

\usepackage[dvips]{graphicx}

\begin{document}

\title{Optimal fast single pulse readout of qubits}

\author{\firstname{Andrey L.} \surname{Pankratov}}
\email{alp@ipm.sci-nnov.ru}
\author{\firstname{Andrey S.} \surname{Gavrilov}}

\affiliation{Institute for Physics of Microstructures of RAS,
GSP-105, Nizhny Novgorod, 603950, Russia}

\begin{abstract}

The computer simulations of the process of single pulse readout from the flux-biased phase qubit is performed in the frame of one-dimensional Schroedinger equation. It has been demonstrated that the readout error can be minimized by choosing the optimal pulse duration and the depth of a potential well, leading to the fidelity of 0.94 for 2ns and 0.965 for 12ns sinusoidal pulses.
\end{abstract}
\date{\today}
 \maketitle
\newpage

In a past decade a serious progress has been achieved in development and creation
of various circuits for quantum computation \cite{qubit},\cite{QIP}. Different sources of decoherence is, however, the main factor, limiting the practical utilization of complex networks of quantum bits \cite{one2}-\cite{shape2}. Recently, it has been demonstrated that, the coherent Rabi oscillations remain nearly unaffected by thermal fluctuations up to temperatures of 1K \cite{temp} (i.e., until the energy of thermal fluctuations $kT$ becomes comparable with energy level spacing $\hbar \omega$ of the qubit), so without degrading already achieved coherence times, phase qubits can be operated at temperatures much higher than those reported till now. This may signal, that relatively large readout errors of practical devices can be attributed not to quantum and thermal fluctuations, but to unoptimal readout of the qubits.  To speed up the readout, the fast single pulse readout (FSPR) technique has been realized and tested \cite{one2}-\cite{one3}. An example of a shallow potential well with the two energy levels $\left|0\right>$ and $\left|1\right>$ is presented in the inset of Fig. 1. The basic idea of the FSPR is the application of a readout pulse in such a way, that during the pulse action the system will tunnel from the state $\left|1\right>$ through the barrier with a probability close to unity, while from the state $\left|0\right>$ it will not tunnel, again, with the probability close to unity. The effect of the shape and duration of the pulse on different error probabilities has been studied in Ref.s \cite{shape1},\cite{shape2}. In particular, the error to excite higher qubit states due to nonadiabaticity of the pulse was analyzed \cite{shape2}. As it has been understood, \cite{one3},\cite{shape2}, the main source of error during qubit readout is due to incomplete discrimination between the two quantum states $\left|0\right>$ and $\left|1\right>$. It has been claimed that the quality of tunneling discrimination depends significantly on the measurement pulse amplitude and this error decreases for longer measurement pulses and with increase of $\Gamma_1/\Gamma_0$ ratio (where $\Gamma_0$ and $\Gamma_1$ are tunneling rates from the states $\left|0\right>$ and $\left|1\right>$, respectively). However, the question about possible compromise between the speed and the fidelity of the readout has not been considered in the literature. The importance of this question is not only due to the reason that one wants both high fidelity and high speed readout, but also due to the fact that any fluctuations will accumulate during longer readout, leading to entangling of states.
\begin{figure}[h]
\resizebox{1\columnwidth}{!}{
\includegraphics{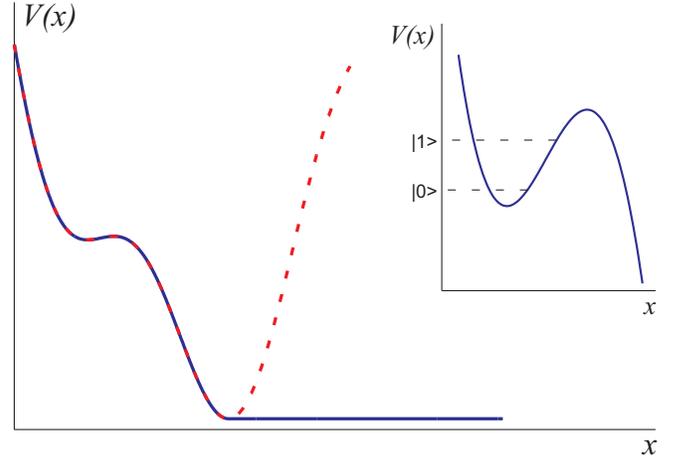}}
{\caption{The profile of a bistable potential. Dashed curve - the original potential, solid curve - the potential with enlarged deep well to simulate the effect of damping. The inset: the enlargement of a shallow potential well.} \label{fig1}}
\end{figure}

Recently, in classical systems subjected to noise and pulsed or periodic
driving, such as Josephson junctions \cite{ALPPRL} and magnetic nanoparticles \cite{ALPPRB}, it has been demonstrated that at a fixed value of the driving amplitude there exists the optimal pulse duration, which minimizes the noise-induced errors. The aim of the present paper is to study the readout error $N$, which is a sum of the two probabilities, $P_{10}$ not to tunnel during the pulse action from the state $\left|1\right>$, and $P_{01}$ to tunnel from the state $\left|0\right>$ (i.e. $N=P_{10}+P_{01}$, while the fidelity $F=1-N$). It is intriguing to understand, would the effect, discovered for classical systems with noise \cite{ALPPRL},\cite{ALPPRB}, be realized in a quantum system, e.g. in a such, as an example of a qubit, described in \cite{one4},\cite{one1},\cite{shape2}. The investigation is performed via computer simulation of the Schroedinger equation and is focused on the dependence of the readout error $N$ versus the pulse amplitude, duration and shape, as well as the depth of a shallow potential well.

\begin{figure}[h]
\resizebox{1\columnwidth}{!}{
\includegraphics{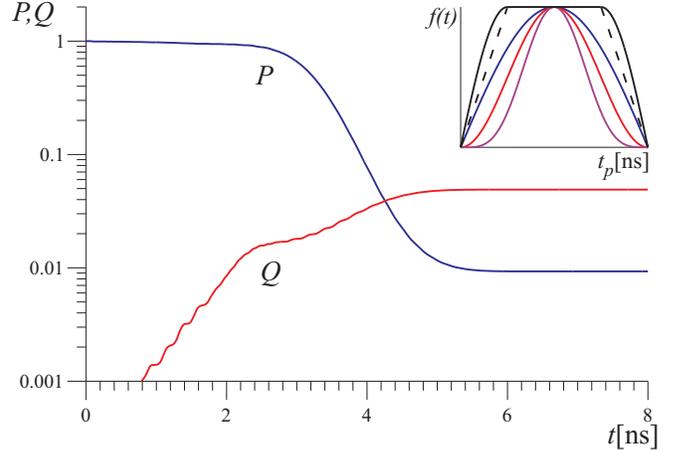}}
{\caption{The evolution of probabilities $P(t)$ and $Q(t)$. Inset: the considered pulses, from top to bottom, sine-trapezoid pulse, trapezoidal pulse, $\sin(\pi t/t_p)$, $\sin^2(\pi t/t_p)$, $\sin^4(\pi t/t_p)$.}
\label{fig2}}
\end{figure}

Let us consider an example of a flux-biased phase qubit \cite{one4},\cite{one1},\cite{shape2}, which is described by the following potential $V(x,t)=E_J\left\{(x-\varphi(t))^2/2\ell-\cos x\right\}$, see Fig. 1, dashed curve. Here $E_J=I_C\hbar/2e$ is the Josephson energy, $x$ is the Josephson phase, $e$ is the electron charge, and $\hbar$ is the Planck constant. We take for the qubit the same parameters as in \cite{one4},\cite{shape2}: the critical current $I_C=1.7\rm{\mu A}$, the inductance of the ring $L=0.72$nH and the capacitance $C=700$fF, thus $\ell=2eI_C L/\hbar=3.71$, ${2e^2}/{\hbar C}=0.6933\times 10^9$Hz, $E_J/\hbar=I_C/2e=5.31\times 10^{12}$Hz, so it is convenient to introduce the "inverse capacitance" $D={2e^2}/{\hbar C}\times 10^{-9}$ and express the time in nanoseconds. The dimensionless external magnetic flux $\varphi(t)$ consists of the two components, the dc component $a_0$, adjusting which the depth of the shallow well can be changed, and the driving readout pulse: $\varphi(t)=2\pi(a_0+Af(t))$, where $A$ is the pulse amplitude, and $f(t)$ is one of $\sin(\pi t/t_p)$, $\sin^2(\pi t/t_p)$, $\sin^4(\pi t/t_p)$, the trapezoid function, which linearly grows and drops for $t\ge t_p/4$ and $t\le 3t_p/4$, and the sine-trapezoid function, which differs from the previous one by the sinusoidal walls, see the inset of Fig. \ref{fig2}. We note that $t_p$ is defined as the full width of the pulse at zero level, not as full width at half maximum. The shift of a potential barrier $a_0=0.81$ is chosen such to allow the six levels to be present in a shallow potential well, see \cite{shape2}. The pulse with the amplitude $A\approx 0.035$ leads to lowering the potential barrier such that only two levels will remain.

The Schroedinger equation for the wave function $\Psi(x,t)$ has the following form:
\begin{equation}
i \frac{\partial\Psi(x,t)}
{\partial t}=-\frac{2e^2}{\hbar C}\frac{\partial^2\Psi(x,t)}{\partial x^2}+\frac{V(x,t)}{\hbar}\Psi(x,t). \label{Sch}
\end{equation}
It is assumed, that the boundary conditions
\begin{equation}
\Psi(c,t)=\Psi(d,t)=0, \label{Bndr}
\end{equation}
at the points $c$ and $d$ are taken far away from the shallow potential well, and do not affect the tunneling process. To prevent from the repopulation error \cite{shape2}, which arises due to the absence of damping in our model, let us introduce an effective damping in the following manner.
Since we are interested in the process of tunneling from the shallow potential well only, we assume that at the bottom of the deep potential well the potential does not grow up, but continues to the right really far away, see Fig. 1, solid curve. In particular, we have taken $c=-3$, $d=797$, while the left minimum
is located at $x_1\approx 1.4$, and the right one at $x_2\approx 6$.
Numerical solution of the Schroedinger equation (\ref{Sch}) with
boundary conditions (\ref{Bndr}) has been performed on
the basis of implicit finite-difference Crank-Nicholson scheme. Typical values of discretization steps are $\triangle x = 0.01$, $\triangle t = 10^{-4}$ns.

In Fig. 2 the evolution of the probabilities $P(t)$ not to tunnel from the state $\left|1\right>$
and $Q(t)$ to tunnel from the state $\left|0\right>$ are presented. In both cases we choose the initial condition (the wave function) to correspond to either $\left|1\right>$ or $\left|0\right>$ stationary state and calculate the survival probabilities $P_{1}(t)$ and $P_{0}(t)$ in the shallow potential well. The probability evolutions $P(t)=P_{1}(t)$ and $Q(t)=1-P_{0}(t)$ are presented in Fig. 2  for the case of sinusoidal pulse with driving amplitude $A=0.034$ and pulse duration $t_p=8$ns. One can see that at the end of the pulse both probabilities are significantly smaller than unity, and our task is to find the optimal parameters of the pulse to minimize the readout error $N=P_{10}+P_{01}=P(t_p)+Q(t_p)$. In difference with \cite{shape2} we do not separate the nonadiabatic error (leading to lifting to higher states), and the error of direct tunneling from the state $\left|0\right>$, because both these errors contribute into the error $P_{01}$, which we calculate. Since for short pulse durations the probability $P_{10}$ must be large, and for long pulses the probability $P_{01}$ increases significantly, there must be an optimal pulse duration, leading to the minimal $N$.
\begin{figure}[h]
\resizebox{1\columnwidth}{!}{
\includegraphics{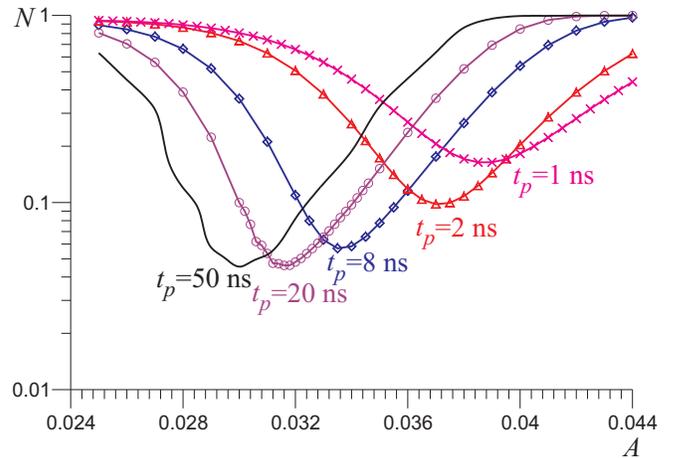}}
{\caption{The readout error $N$ versus the pulse amplitude $A$ for the pulse $\sin(\pi t/t_p)$.}
\label{fig3}}
\end{figure}

The readout error $N$ versus the pulse amplitude $A$ for the sinusoidal pulse is presented in Fig. 3 for different values of the pulse width $t_p$. It is seen that $N$ is very sensitive to the pulse amplitude, especially for large pulse durations, and that the minimal value of $N$ decreases with increase of $t_p$. Similar figure for the only one pulse width has been presented in Ref. \cite{one3}. However, from Fig. \ref{fig3} one can see that for the pulse durations $t_p\ge 8$ns the minimal value of $N$ is almost the same, while below $t_p\le 2$ns $N$ increases significantly.
\begin{figure}[h]
\resizebox{1\columnwidth}{!}{
\includegraphics{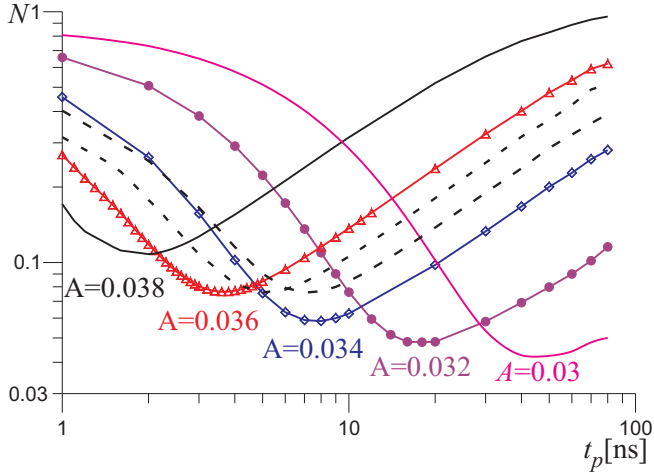}}
{\caption{The readout error $N$ versus the pulse width $t_p$ for the pulse $\sin(\pi t/t_p)$. The dashed curves correspond to $\sin^2(\pi t/t_p)$, $\sin^4(\pi t/t_p)$ pulses with $A=0.036$ (from left to right).}
\label{fig4}}
\end{figure}

The readout error $N$ versus the pulse width $t_p$ for the sinusoidal pulse is presented in Fig. \ref{fig4} for different values of the pulse amplitude $A$. One can see that $N$ has strongly pronounced minimum, and with increase of $A$ this minimum shifts to smaller tunneling times, correspondingly speeding up the readout. With increase of $A$ the value of the error $N$ at the minimum increases, so one should find the compromise between the error and the speed. However, in the range of small amplitudes $N$ increases insignificantly, compare the curves for $A=0.03$, $A=0.032$ and $A=0.034$, and for the considered parameters the amplitude $A=0.034$ can be chosen as a compromise value, leading to the fast readout with the high fidelity $F=0.94$. The results, presented in Fig.s \ref{fig3} and \ref{fig4}, on one hand, demonstrate similar dependence of the readout error vs pulse width as for classical systems \cite{ALPPRL},\cite{ALPPRB}. On the other hand, the dependence on the driving amplitude also has a minimum, which outlines the quantum nature of the described phenomena, while for a classical system the error monotonously decreases with increase of the pulse amplitude. Another important deviation from a classical system is the impossibility to use the rectangular readout pulses (which in a classical case leads to the minimal noise-induced errors). In the qubit the rectangular pulse leads to nonadiabaticity of the tunneling event that results to lifting to higher eigenstates and considerable increase of the probability to tunnel from $\left |0\right>$ state and thus to much larger values of $N$ than for all other considered pulse shapes.

\begin{figure}[h]
\resizebox{1\columnwidth}{!}{
\includegraphics{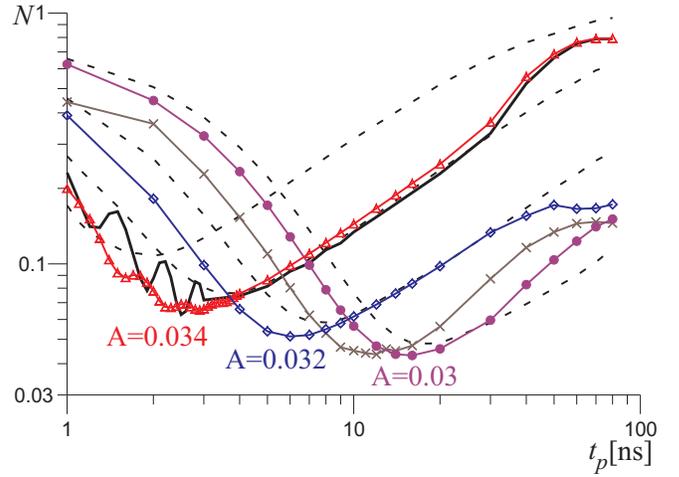}}
\caption{The readout error $N$ versus the pulse width $t_p$ for the trapezoid pulse with $A=0.034$ (thick solid curve) and sine-trapezoid pulses (solid curves with triangles, diamonds and circles for amplitudes $A=0.034;0.032;0.03$). The curves for the pulse shape $\sin(\pi t/t_p)$ from Fig. \ref{fig4} with $A=0.038;0.036;0.034;0.032$ are given by dashed curves for comparison. Solid curve with crosses - long sine-trapezoid pulse for $A=0.03$.}
\label{fig6}
\end{figure}
The readout error for different pulse shapes $\sin^2(\pi t/t_p)$ and $\sin^4(\pi t/t_p)$, $A=0.036$ is presented in Fig. \ref{fig4} by dashed curves. As one can see, the minimal value of the error is pretty much the same, but the optimal readout time is significantly smaller for $\sin(\pi t/t_p)$ just due to the fact, that the top of the pulse in this case is more flat.
In accordance with these results the trapezoid pulse should lead to smaller $N$. This is indeed so, as one may see from Fig. \ref{fig6}, thick solid curve. However, the dependence of $N$ vs $t_p$ demonstrates oscillations, and it is not easy to properly choose the optimal pulse width. Better situation can be achieved, if one substitutes the linear walls in the trapezoid by $\sin(2\pi t/t_p)$, see solid curves with triangles, diamonds and circles for amplitudes $A=0.034$, $A=0.032$ and $A=0.03$, respectively.
In comparison with pure sinusoidal pulse, $\sin(\pi t/t_p)$, one can gain about 15-20 $\%$ in the readout error, and, simultaneously, up to 50 $\%$ in the readout speed, compare with the dashed curves, taken from Fig. \ref{fig4} for $A=0.038$, $A=0.036$, $A=0.034$, $A=0.032$, respectively. Further increase of the readout speed can be achieved, if one increases the duration of the flat part of the sine-trapezoid pulse from $t_p/8$ to $7t_p/8$, see solid curve with crosses for $A=0.03$. Transferring to the limit of a rectangular pulse, one should stop somewhere, since further increase of the flat part of the pulse will give little increase in speed, but will obviously lead to the increase of nonadiabatic error \cite{shape2}.
\begin{figure}[h]
\resizebox{1\columnwidth}{!}{
\includegraphics{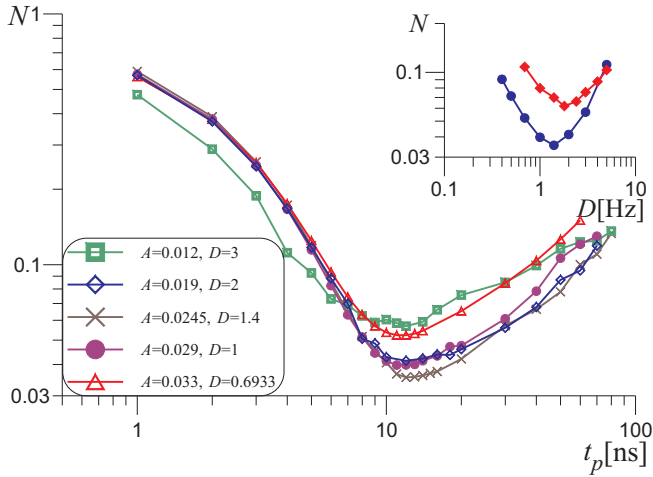}}
\caption{The readout error $N$ versus the pulse width $t_p$ for the pulse $\sin(\pi t/t_p)$ with different amplitudes and inverse capacitances $D$. Inset: the readout error $N(D)$ for the pulse duration $t_p=12$ns (circles) and $t_p=2$ns (diamonds).}
\label{fig7}
\end{figure}

Finally, let us briefly demonstrate, how adjusting the depth of a shallow potential well, the readout error can be decreased further. If this well is deep enough, there are many levels inside, and while both $\Gamma_0$ and $\Gamma_1$ are small, their values are close to each other, which complicates the discrimination between the states $\left|0\right>$ and $\left|1\right>$. If the well is too shallow, the tunneling from the state $\left|1\right>$ may occur even without the driving pulse, and the discrimination between the states, again, will be pure. Therefore, there must be the optimal depth of the well, leading to the minimal readout error. The depth of the well can be adjusted either by variation of the constant magnetic field component $a_0$, or by variation of the capacitance $C$ of the Josephson junction. The latter case is illustrated in Fig. \ref{fig7}, where the readout error $N$ is presented for different amplitudes and values of the inverse capacitance $D={2e^2}/{\hbar C}\times 10^{-9}$ in such a way that the minimum of $N$ remains in approximately the same region of pulse duration $t_p\approx 12$ns. Fixing $t_p$, from the main part of Fig. \ref{fig7} one can extract the values of $N$ and plot them as a function of $D$, as it is done in the inset of Fig. \ref{fig7}, the curve with circles. The readout error $N(D)$ demonstrates pronounced minimum and choosing the optimal value of $D\approx 1.4$Hz one can decrease the value of $N$ down to 0.0355, leading to the fidelity $F=0.965$  (note that, $N=0.0523$ for $D=0.6933$Hz, $A=0.033$). The same procedure can be performed for $t_p=2$ns (see the curve with diamonds in the inset): while for $D=0.6933$Hz the fidelity was about 0.9, at the minimum, which is shifted to $D\approx 1.8$Hz, $F\approx 0.94$. The location of the minimum of $N(D)$ corresponds to something between three or two energy levels inside the shallow potential well. We note, that rather fast readout of 2ns duration with acceptably high fidelity of 94\% is achieved with a simple sinusoidal pulse, whose generation does not require complex pulse shaping hardware \cite{PSH}. For 2ns sine-trapezoid pulse the fidelity can be increased to 0.947. We, therefore, believe that the proper engineering of the qubit, together with the optimal adjustment of the readout pulse duration, will finally lead to creation of high fidelity and high-speed readout qubits.

In conclusion, we have performed the computer simulations of the process of pulsed readout from the flux-biased phase qubit within the model of one-dimensional time-dependent Schroedinger equation. It has been demonstrated that by choosing the optimal pulse duration the readout error can be minimized. Further decrease of the readout error can be achieved by variation of the depth of the shallow potential well.

The authors wish to thank M.A. Silaev for discussions and useful comments.
The work has been supported by RFBR (projects 09-02-00491 and 08-02-97033).

\end{document}